\documentclass[12pt]{article}

\usepackage{preprint}
\usepackage{amssymb}
\usepackage{amsmath}
\usepackage{wasysym}  
\usepackage{graphicx}
\usepackage{changebar}

\begin{document}

\hyphenation{cal-o-rim-e-ter}

\title{A new evaluation of the pion weak form factors}

\author{\normalsize D.~Po\v cani\'c, (for the PIBETA collaboration) \\[1ex]
\normalsize
 Physics Department,
 University of Virginia,
 Charlottesville, VA~22904-4714, USA 
               }
\date{\normalsize \vspace*{-3ex}}
\maketitle
\thispagestyle{empty}\pagestyle{plain}

\begin{abstract}

We have used the PIBETA large acceptance detector for a precise
measurement of the $\pi^+ \to e^+\nu\gamma$ radiative decay at rest,
with broad phase space coverage.  Using the CVC value for the pion
vector form factor, $F_V = 0.0259(5)$, we have obtained a new value of
the pion axial-vector form factor $\gamma = F_A/F_V = 0.429(14)$.
However, significant deviations from the SM predictions are evident in
our data.  The discrepancy can be accounted for by introducing a pion
tensor form factor of $F_T = -0.0016(3)$.  The question of the true
nature of the observed deviations remains open pending further
theoretical and experimental study.

\end{abstract}

\section{Motivation}

The PIBETA experiment\cite{pibeta} at the Paul Scherrer Institute
(PSI) is a comprehensive set of precision measurements of the rare
decays of the pion and muon.  The goals of the experiment's first
phase are:

\begin{itemize}

\item[(a)] To improve the experimental precision of the pion beta
branching ratio ($BR$), $\pi^+ \to \pi^0 e^+ \nu$ (referred to as
$\pi_{e3}$, or $\pi\beta$), from the present $\sim 4\,\%$ to $\sim
0.5\,\%$.

\item[(b)] To measure the branching ratio of the radiative pion decay
$\pi\to e\nu\gamma$ ($\pi_{e2\gamma}$, or RPD), enabling a precise
evaluation of the pion axial-vector form factor $F_A$ and tensor form
factor $F_T$, predicted to vanish in the Standard Model (SM).

\item[(c)] An extensive measurement of the radiative muon decay rate,
$\mu\to e \nu \bar{\nu} \gamma$, with broad phase space coverage,
enabling a search for non-\,(V$-$A) admixtures in the weak Lagrangian.

\end{itemize}

The experiment's second phase calls for a precise measurement of the
$\pi\to e \nu$ (known as $\pi_{e2}$) decay rate, used for $BR$
normalization in the first phase.  The current $\sim 0.35\,\%$
accuracy would be improved to under 0.2\,\%, in order to provide a
precise test of lepton universality, and, hence, of certain extensions
to the Standard Model.

In this report we focus on part (b) above.  Radiative pion decay
offers unparalleled access to information on the pion's structure.
Given the unique role of the pion as the quasi-Goldstone boson of the
strong interaction, the implications are far reaching.

In the Standard Model description \cite{Bry82} of the radiative pion
decay $\pi^+\to e^+\nu\gamma$, where $\gamma$ is a real or virtual
photon ($e^+e^-$ pair), the decay amplitude ${\cal M}$ depends on the
vector V and axial vector A weak hadronic currents. Both currents give
rise to structure-dependent terms SD$_\text{V}$ and SD$_\text{A}$
associated with virtual hadronic states, while the axial-vector
current alone causes inner bremsstrahlung process IB from the pion and
positron.

The IB contribution to the decay probability can be calculated in a
straightforward manner using QED methods. The structure-dependent
amplitude is parameterized by the vector form factor $F_V$ and the
axial vector form factor $F_A$ that have to be extracted from
experiments.

The statistics and overall accuracy of the present experimental data
on the radiative pion
decay~\cite{Dep63p,Ste74,Ste78,Bay86,Pii86,Bol90} cannot rule out
contributions from other allowed terms in the interaction lagrangian,
namely the scalar S, pseudoscalar P, and tensor T admixtures
\cite{Mur85}.  Nonzero values of any of these amplitudes would imply
new physics outside the Standard Model.  In particular, reports from
the ISTRA collaboration \cite{Pob90,Pob03} have indicated a nonzero
tensor term, $F_T = -0.0056\,(17)$.  A careful analysis by Herczeg of
the existing beta decay data set could not rule out such a value of
$F_T$, presumed to be due to leptoquarks \cite{Her94}.  A new
intermediate chiral boson with an anomalous interaction with matter
has been proposed by Chizhov in order to account for the apparent
non-(V$-$A) behavior in RPD \cite{Chi93}.

Moreover, the ratio of $F_A/F_V$ in $\pi\to e\nu\gamma$ decay directly
determines the chiral perturbation theory parameter sum
$(l_9+l_{10})$, or, equivalently, $\alpha_E$, the pion polarizability
\cite{Hol90,Bij97,Gen03}.  These quantities are of longstanding
interest since they are among the few unambiguous predictions of
chiral symmetry and QCD at low energies.  The current status of these
measurements is not satisfactory, as there is considerable scatter
among the various experimental determinations of $F_A/F_V$, and the
accepted PDG average value has a 14\,\% uncertainty.

\section{Experimental method}

\begin{figure}[ht]
\noindent\hbox to \textwidth{\hfill
 \resizebox{0.95\textwidth}{!}
            {\includegraphics{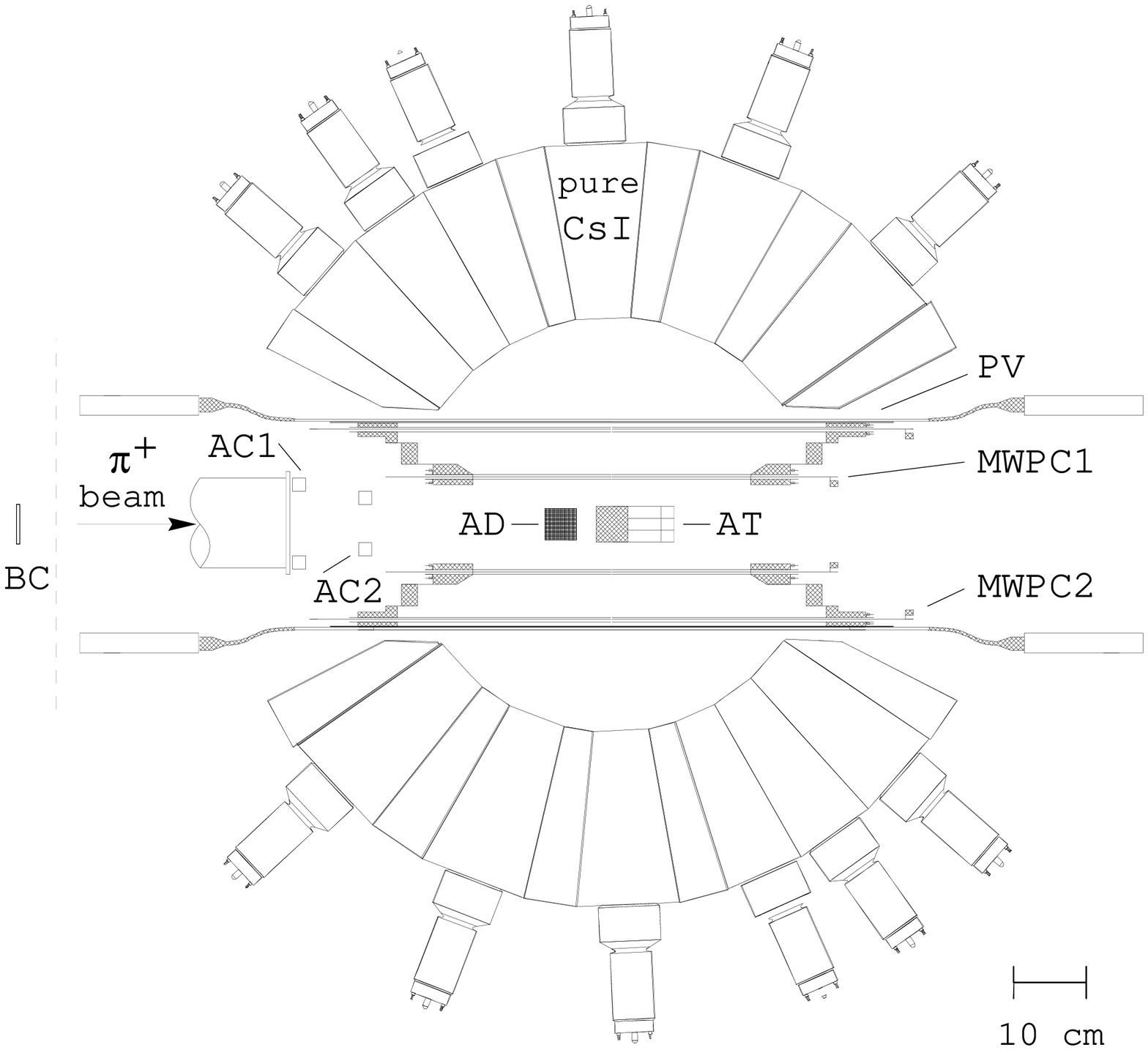}} \hfill}
\noindent
\begin{minipage}[b]{0.45\textwidth}
 \caption{(a) [above] Schematic cross section of the PIBETA apparatus
   showing the main components: beam entry counters (BC, AC1, AC2),
   active degrader (AD), active target (AT), wire chambers (MWPCs) and
   support, plastic veto (PV) detectors and PMTs, pure CsI calorimeter
   and PMTs.  (b) [right] Axial (beam) view of the central detector
   region showing the 9-element active target and the charged particle
   tracking detectors. \vspace*{3ex} \label{fig1}}
\end{minipage}
\hfil
\resizebox{0.5\textwidth}{!}
             {\includegraphics{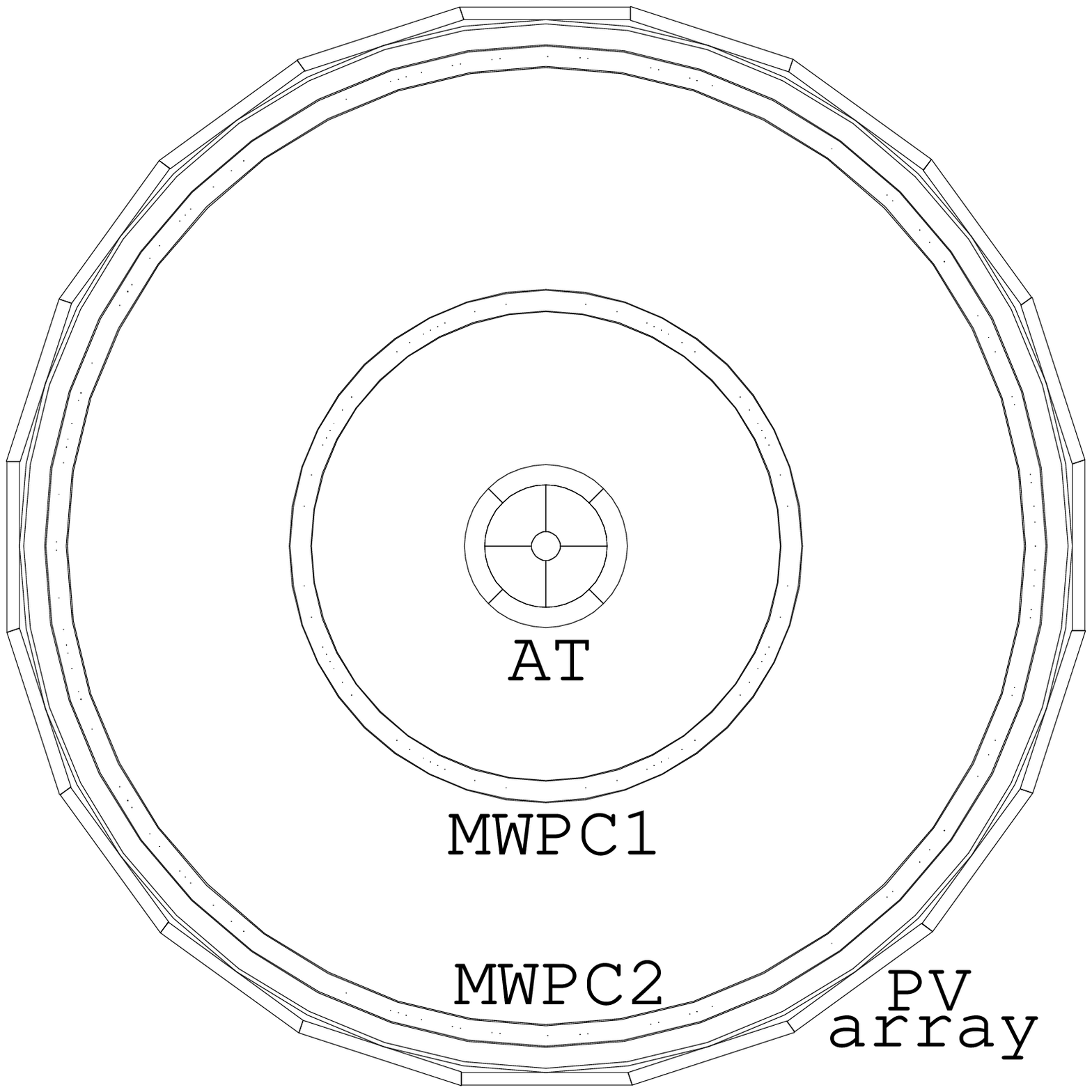}}
\end{figure}

The PIBETA apparatus is a large solid angle non-magnetic detector
optimized for detection of photons and electrons in the energy range
of 5--150$\,$MeV with high efficiency, energy resolution and solid
angle.  The main sensitive components of the apparatus, shown and
labeled in Fig.~\ref{fig1}, are: 

\begin{itemize} 

\item[(a)] beam definining plastic scintillator detectors: BC, a thin
forward beam counter, AC$_1$ and AC$_2$, cylindrical active
collimators, AD, an active degrader, AT, a 9-element segmented active
target;
\item[(b)] charged particle tracking and id: MWPC$_1$ and MWPC$_2$,
cylindrical chambers, and PV, a 20-bar segmented thin plastic
scintillator hodoscope;
\item[(c)] a 240-element segmented spherical pure-CsI shower
calorimeter, subtending a solid angle of $\sim 80\,$\% of $4\pi$.
\end{itemize}

\bigskip

\clearpage

The building and testing of the detector components were completed in
1998, followed by the assembly and commissioning of the full detector
apparatus.  Data acquisition with the PIBETA detector started in the
second half of 1999, initially at a reduced pion stopping rate, as
planned.  The experiment ran subsequently during 2000 and 2001 at
$\sim 1\,$MHz $\pi^+$ stopping rate.

A full complement of twelve fast analog triggers comprising all
relevant logic combinations of one- or two-arm, low- or high
calorimeter threshold , prompt and delayed (with respect to $\pi^+$
stop time), as well as a random and a three-arm trigger, were
implemented in order to obtain maximally comprehensive and unbiased
data samples.

\section{Results and analysis}

\begin{figure}[t]
\newlength\widd
\setlength\widd{0.48\textwidth}
\resizebox{\widd}{!}{\includegraphics{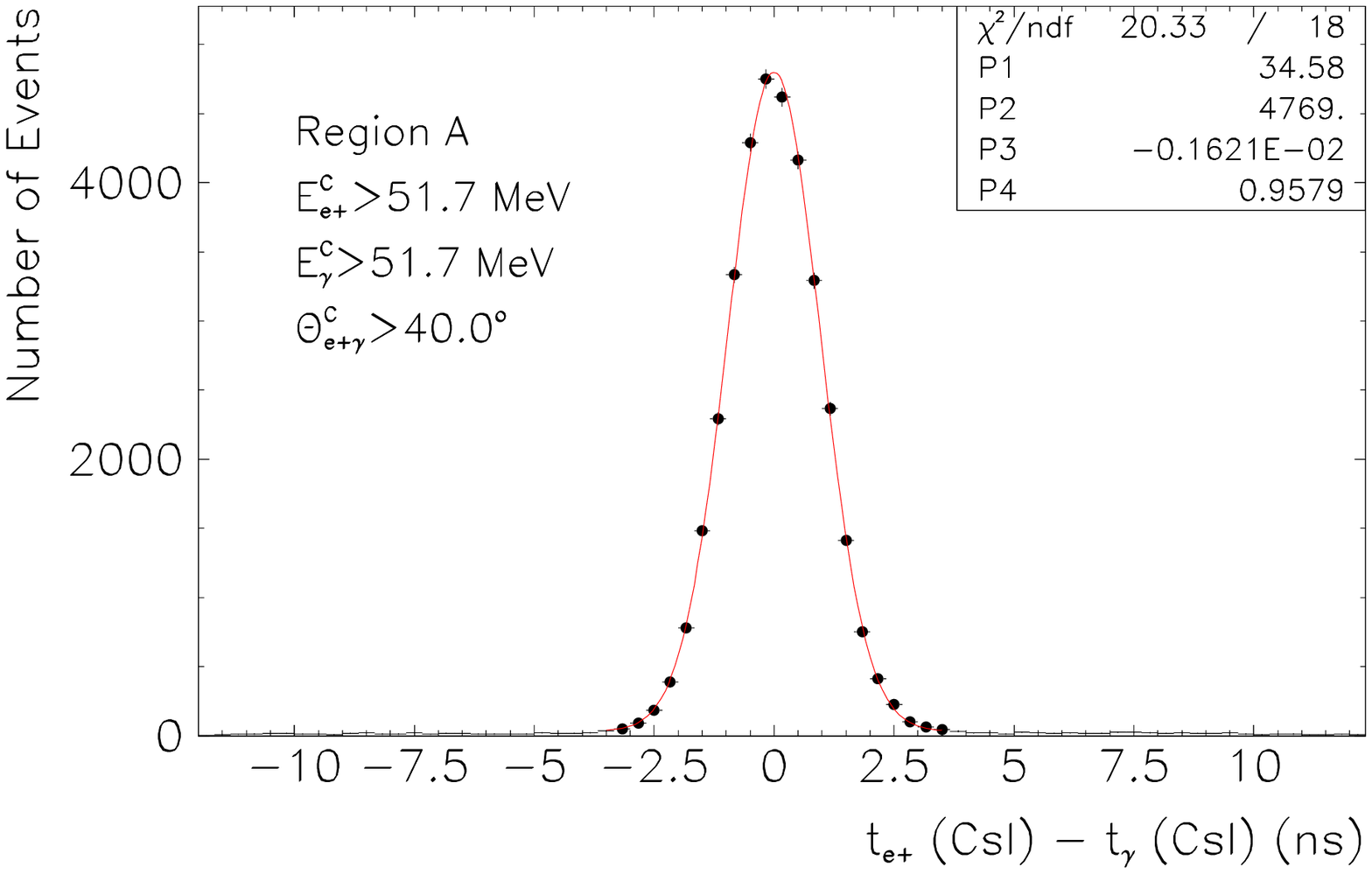}}  \hfil
\resizebox{\widd}{!}{\includegraphics{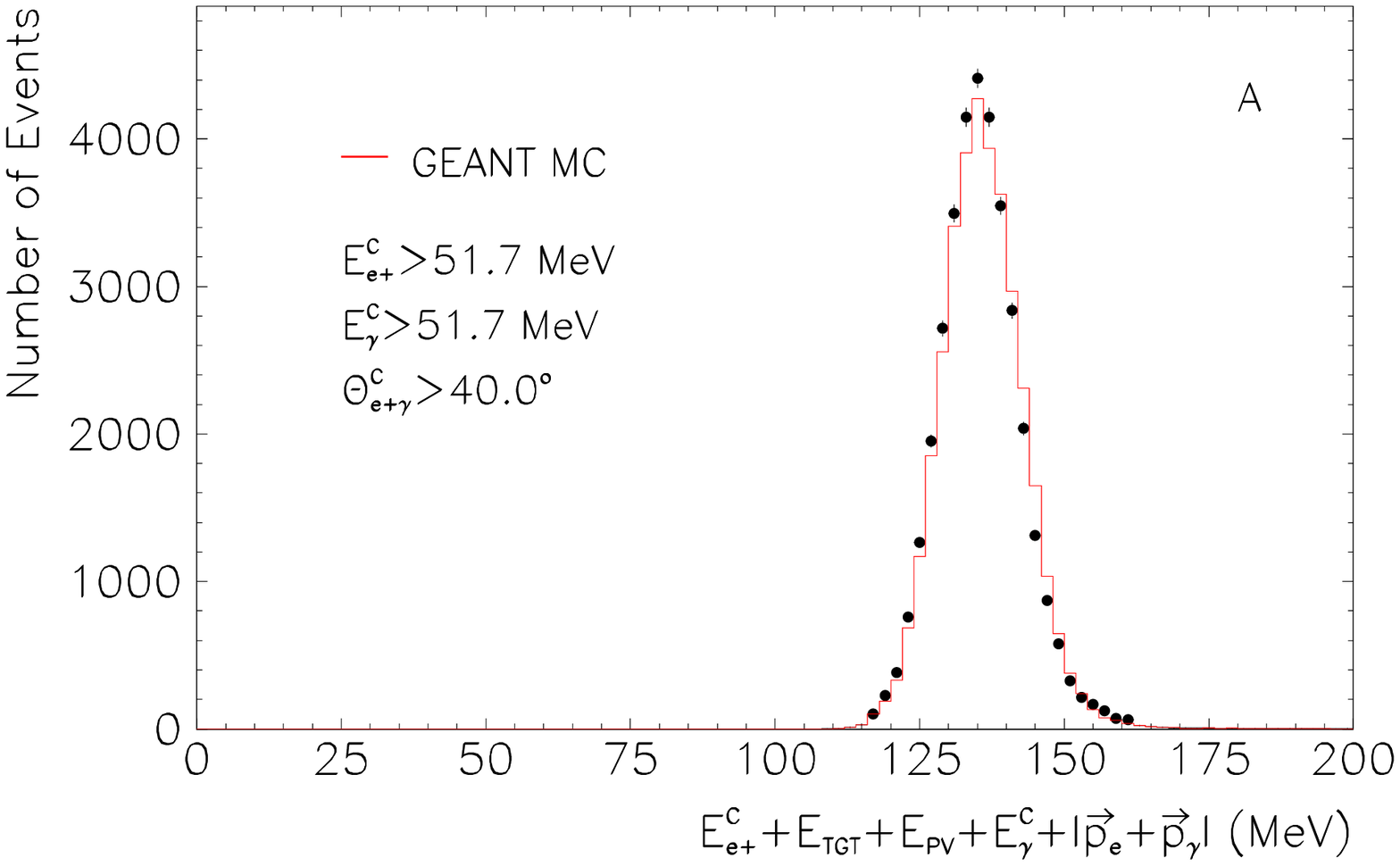}}\\
\resizebox{\widd}{!}{\includegraphics{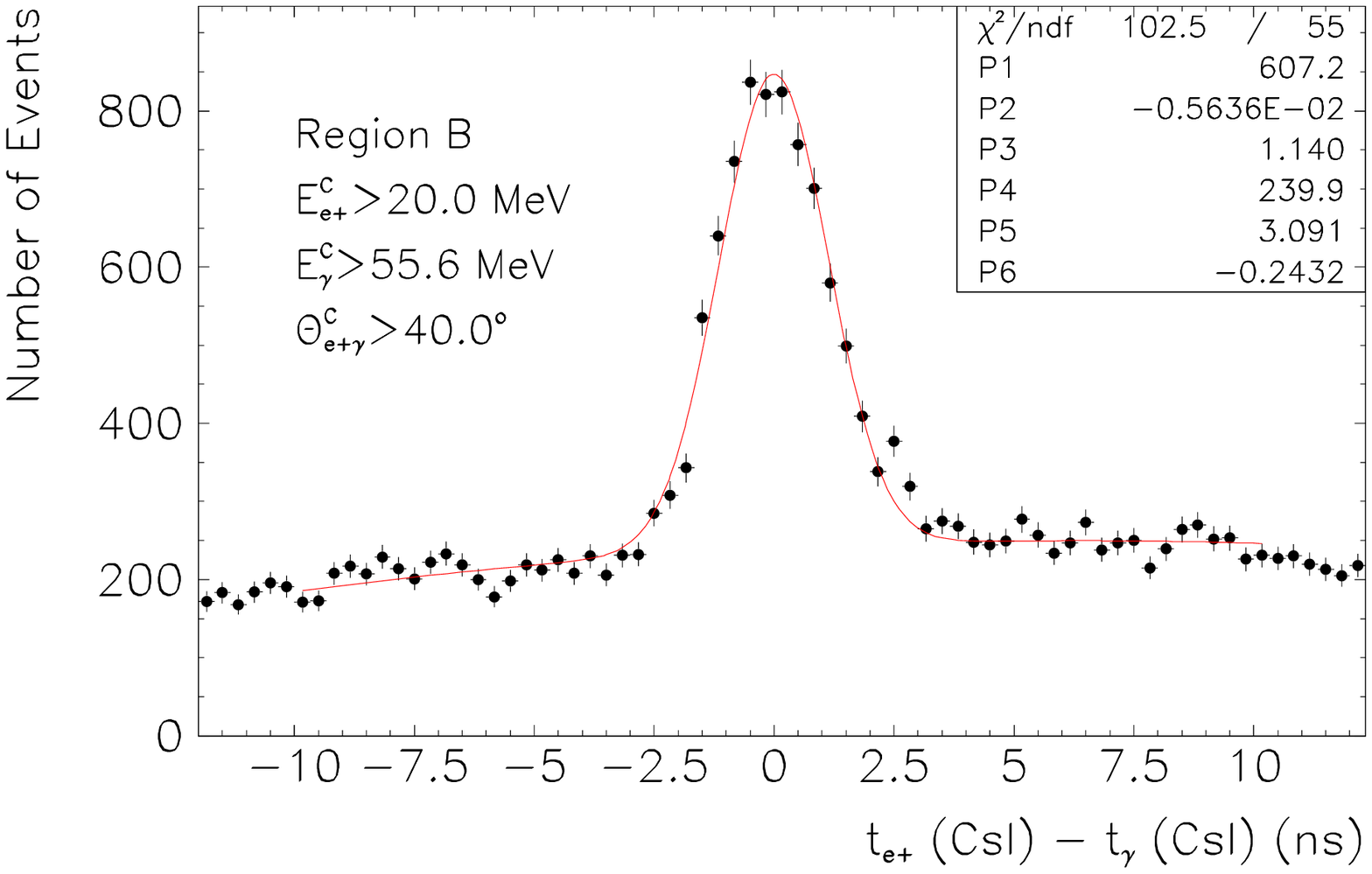}}  \hfil
\resizebox{\widd}{!}{\includegraphics{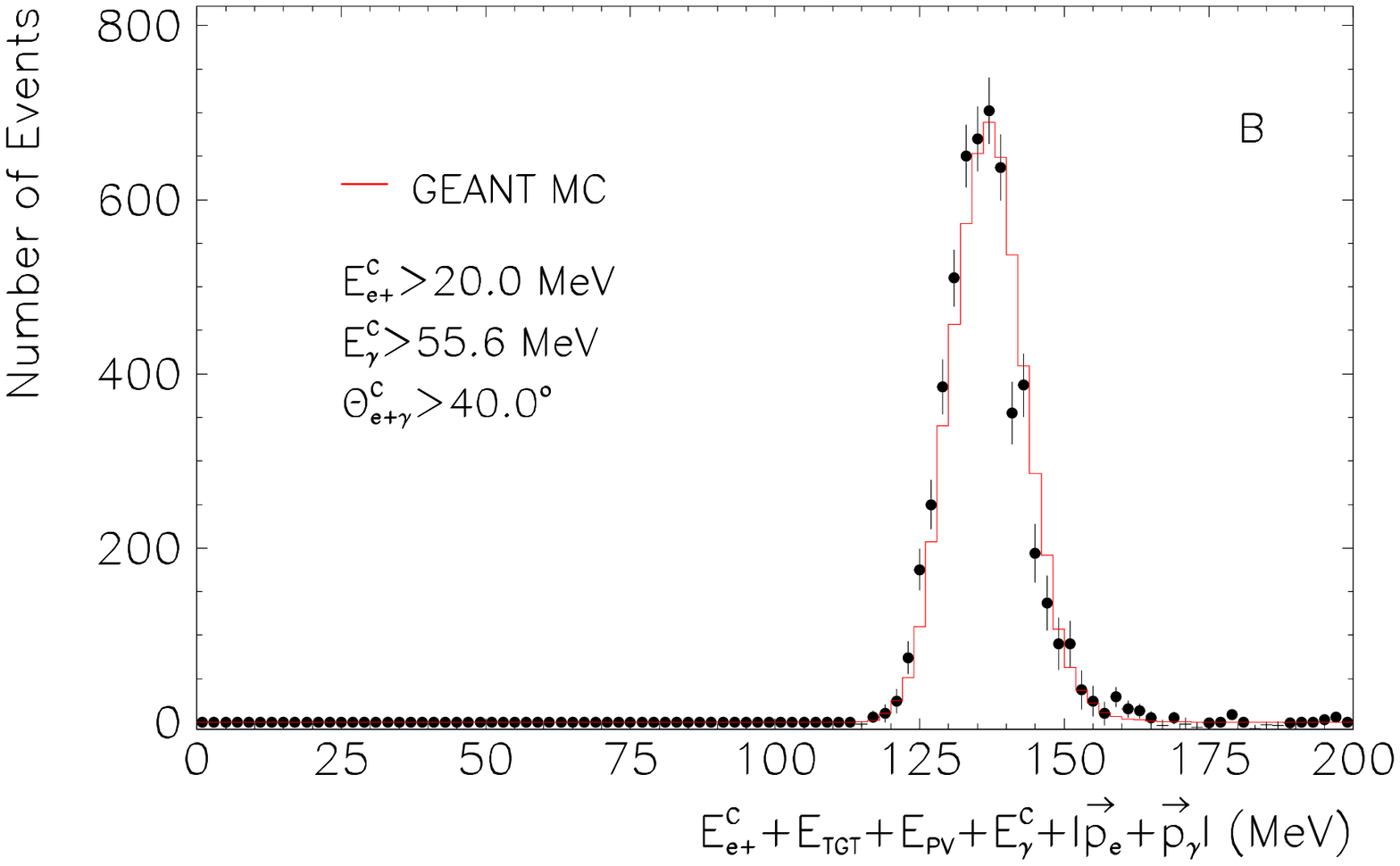}}\\
\resizebox{\widd}{!}{\includegraphics{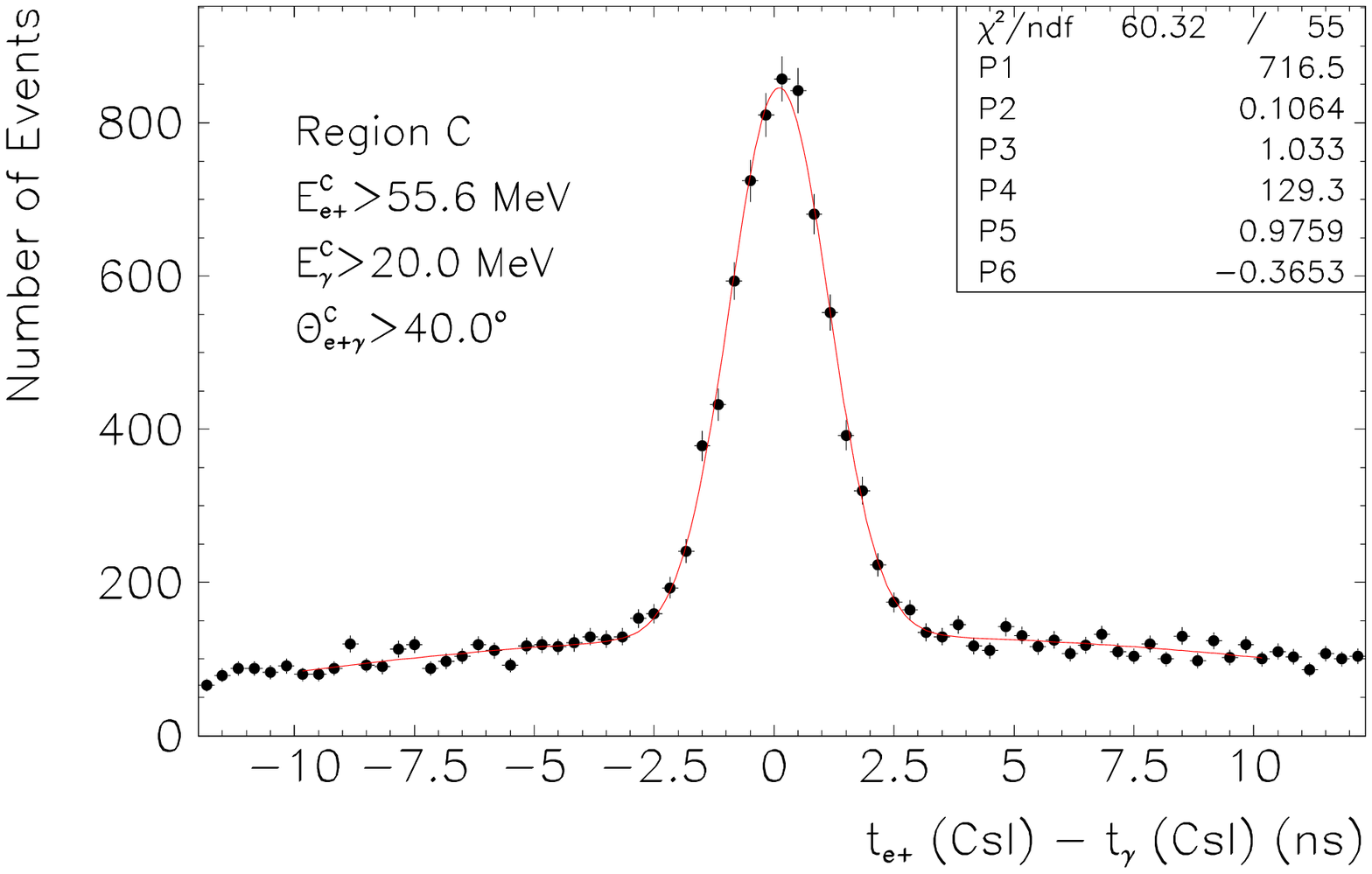}}  \hfil
\resizebox{\widd}{!}{\includegraphics{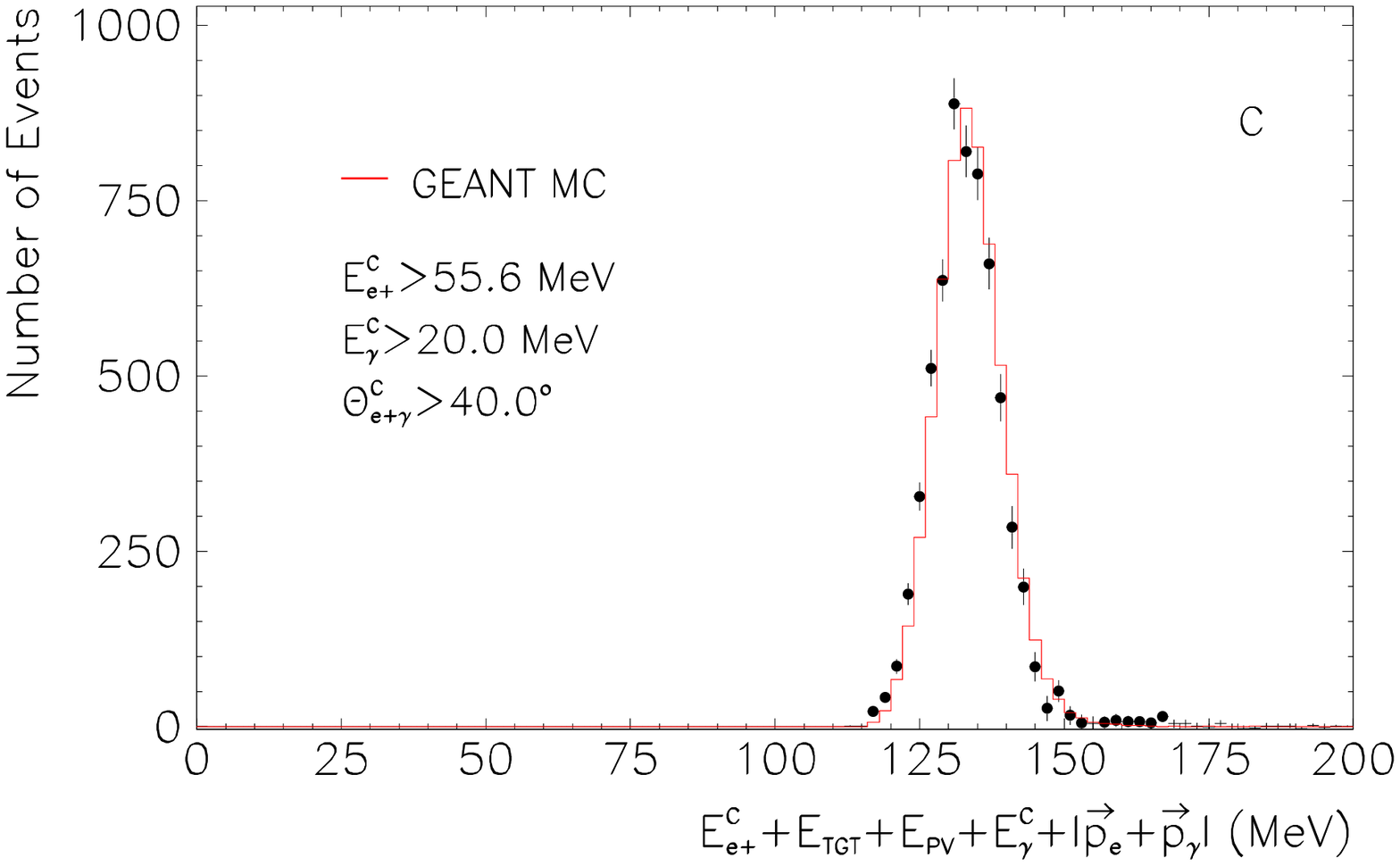}}

\caption{$e$-$\gamma$ timing difference for $\pi\to
e\nu\gamma$ decay events in regions A, B, and C (left panels, top to
bottom, respectively).  The right panels plot the reconstructed
invariant $\pi^+$ mass variable, for our radiative pion decay event
data (dots) in regions A, B, and C, and simulation (histogram).}
\label{fig:pienug:t:mm}

\end{figure}

We have used both one- and two-arm triggers to collect a clean set of
$>43\,$k $\pi^+\to e^+\nu\gamma$ decays.  The PIBETA experiment has
good sensitivity in three distinct regions in the RPD phase space
\begin{itemize}

\item[A:] with $e^+$ and $\gamma$ emitted into opposite hemispheres,
each with energy exceeding that of the Michel edge ($E_M \simeq
52\,$MeV), recorded in a two-arm trigger,

\item[B:] with an energetic photon ($E_\gamma > E_M$), and $E_{e+}
\geqslant 20\,$MeV, recorded in the one-arm trigger,

\item[C:] with an energetic positron ($E_{e+} > E_M$), and $E_\gamma
\geqslant 20\,$MeV, recorded in the one-arm trigger.

\end{itemize}
The quality of our $\pi^+\to e^+\nu\gamma$ data is illustrated in
Fig.~\ref{fig:pienug:t:mm}.  The left panel of the figure shows the
timing difference between the detected $e^+$'s and $\gamma$'s for the
three kinematic regions of the decay, as noted.  Plotted in the right
panel of Fig.~\ref{fig:pienug:t:mm} is the kinematic variable $E_{e^+}
+ E_\gamma + |\vec{p}_e + \vec{p}_\gamma|$ (``invariant pion mass''),
with the out-of-time events subtracted; it is in excellent agreement
with Monte Carlo simulations, demonstrating that our RPD signal is
clean in all three regions.

Together, the three regions overconstrain the Standard Model
parameters describing the decay, and thus allow us to examine possible
new information about the pion's hadronic structure, or non-(V$-$A)
interactions.  Details of the full data analysis require a longer
presentation than is possible here, so we only give a brief
discussion.

Simultaneous as well as separate fits of our data in regions A, B and
C confirm the theoretically expected ratio of $F_A/F_V \simeq 0.5$.
In particular, we find in a Standard Model fit with $F_A$ set free,
and $F_V= 0.0259(5)$ fixed by the CVC hypothesis
\begin{equation}
   \gamma = \frac{F_A}{F_V} = 0.429\pm 0.014 \qquad {\rm or}\qquad
   F_A = 0.01111(36)\ ,                \label{eq:fa-fv}
\end{equation}
which represents a fourfold improvement in accuracy over the current
world average.  This result provides the most accurate experimental
determination to date of the pion polarizability, $\alpha_E$, and of
the $\chi PT$ parameter sum $(l_9+l_{10})$.

\clearpage

\begin{figure}[bh]
\begin{center}
\includegraphics[width=0.8\textwidth]{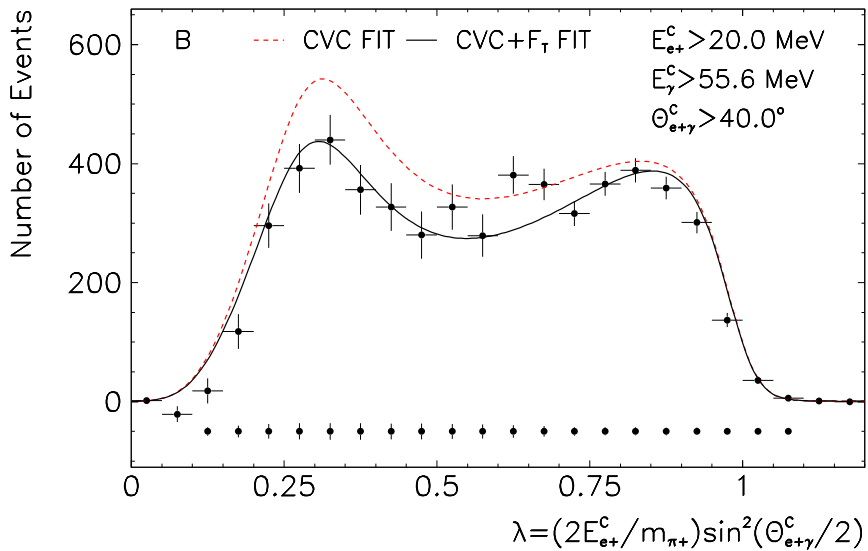}
\end{center}
\caption{Measured spectrum of $\lambda =$ $(2E_e/m_\pi)
\sin^2(\theta_{e\gamma}/2)$ in 
RPD for the kinematic region B, with limits noted in the figure.
Dashed curve: fit with the pion form factor $F_V = 0.0259$ fixed by
the CVC hypothesis, $F_T=0$, and $F_A$ free.  Solid curve: $F_V$ and
$F_A=0.01111$ from the first fit, this time with $F_T$ released to vary
freely, resulting in $F_T = -0.0016\,(3)$.  Error bars on the points
at bottom of graph reflect the expected uncertainties in the proposed
dedicated measurement of the RPD.  \label{fig:gamma:B}}

\end{figure}

However, our simultaneous fits demonstrate a statistically significant
deficit of RPD yield in region B, compared to V$-$A calculations with
the above values of $F_V$ and $F_A$.  In order to account for this
deficit we need to introduce a tensor term destructively interfering
with the IB amplitude, which has led to the result 
\begin{equation}
     F_T = -0.0016\,(3) \ .              \label{eq:ft}
\end{equation}
The effect of the tensor term is best illustrated in
Fig.~\ref{fig:gamma:B}. showing the measured distribution of $\lambda$,
a relevant kinematic variable based on $E_e$ and $\theta_{e\gamma}$,
in region B.  The solid curve, including a negative tensor form
factor, is clearly preferred by the data, which is reflected in much
reduced value of $\chi^2$.

Unfortunately, region B is characterized by the lowest
signal/accidental background (S/B) ratio of the three regions, cf.\
Fig.~\ref{fig:pienug:t:mm}, and, consequently the data in region B are
of lower quality than in region C, and especially region A.  This is a
consequence of the priority having been given to $\pi\beta$ data in
PIBETA's first phase which favored higher rate running.  Even with
this compromise our data unambiguously deviate from the SM (V$-$A)
behavior by over $5\sigma$.  At the same time, though, they rule out
the large ISTRA value of $(-F_T) \simeq 0.006 - 0.012$
\cite{Pob90,Pob03}.

Given the unique sensitivity of region B to the putative tensor
interaction, the longstanding interest in this open question, and the
less than optimal quality of our data in region B, we plan to revisit
RPD with a dedicated run optimized for regions B and C.  The breakdown
of the counting statistics in the first-phase RPD data set is $N_A :
(N_B + N_C) =\rm 31\,k : 12\,k$.  Running for three months with a
reduced pion stopping rate and suitably modified trigger, we can
collect up to 20\,k clean events with S/B $>40:1$ in regions B and C.
The expected error limits for these planned data in region B are shown
in Fig.~\ref{fig:gamma:B} on the line of data points at the bottom of
the plot.  In the process we will also improve our precision on the
$F_A/F_V$ ratio.

We note in closing that the discrepancy from the SM behavior observed
in our RPD data must also be reexamined from the theoretical side.  It
is imperative to ascertain that all known SM processes are correctly
included in the published theoretical amplitudes \cite{Bry82,Nik91}
used in our analysis.

This material is based upon work supported by the National Science
Foundation under Grant No.\ 0098758.


\end{document}